\documentclass[10pt,twocolumn,letterpaper]{article}

\usepackage{iccv}              

\usepackage{soul}      
\usepackage{dirtytalk} 
\usepackage{booktabs}  
\usepackage{tabularx}  
\usepackage{colortbl}  
\usepackage{xcolor}    
\usepackage{array}     
\usepackage{graphicx}
\usepackage{cuted}
\usepackage{capt-of}
\usepackage{stfloats}

\definecolor{iccvblue}{rgb}{0.21,0.49,0.74}
\usepackage[pagebackref,breaklinks,colorlinks,allcolors=iccvblue]{hyperref}

\newcommand{\sysname}{Geo-Visual Agent\xspace}   
\newcommand{\sysnames}{Geo-Visual Agents\xspace}   
\newcommand{\myquote}[1]{{\textit{\say{#1}}}}    

\def\paperID{*****} 
\def\confName{ICCV}
\def\confYear{2025}

\title{``Does the cafe entrance look accessible? Where is the door?'' \\ 
Towards Geospatial AI Agents for Visual Inquiries\vspace{-1.0em}}

\author{
    Jon E. Froehlich\textsuperscript{1,2} \quad
    Jared Hwang\textsuperscript{1} \quad
    Zeyu Wang\textsuperscript{1} \quad
    John S. O'Meara\textsuperscript{1} \quad
    Xia Su\textsuperscript{1} \quad
    William Huang\textsuperscript{3} \\
    Yang Zhang\textsuperscript{3} \quad
    Alex Fiannaca\textsuperscript{4} \quad
    Philip Nelson\textsuperscript{2} \quad
    Shaun Kane\textsuperscript{2}
    \vspace{0.3em} \\
    \textsuperscript{1}University of Washington \quad
    \textsuperscript{2}Google Research \quad
    \textsuperscript{3}UCLA
    \textsuperscript{4}Google DeepMind
    \vspace{0.3em} \\
    {\tt\small jonf@cs.uw.edu}
}

\begin{document}
\maketitle

\begin{strip}
  \vspace{-4.4em} 
  \centering
  \includegraphics[alt={A screenshot of the StreetViewAI system showing the augmented Google Street View interface for a AI-generated description that reads "You are facing a multi-storying brick building with the words 'Irving Farm New York` written boldly above the entrance. To your left, there are some tables and chairs. There is a sign for Thompson Street on the right." Below the description, the user has also engaged in an interactive AI chat asking questions like "Is there a pedestrian light here?" and "What does the entrance look like?"}, width=\textwidth]{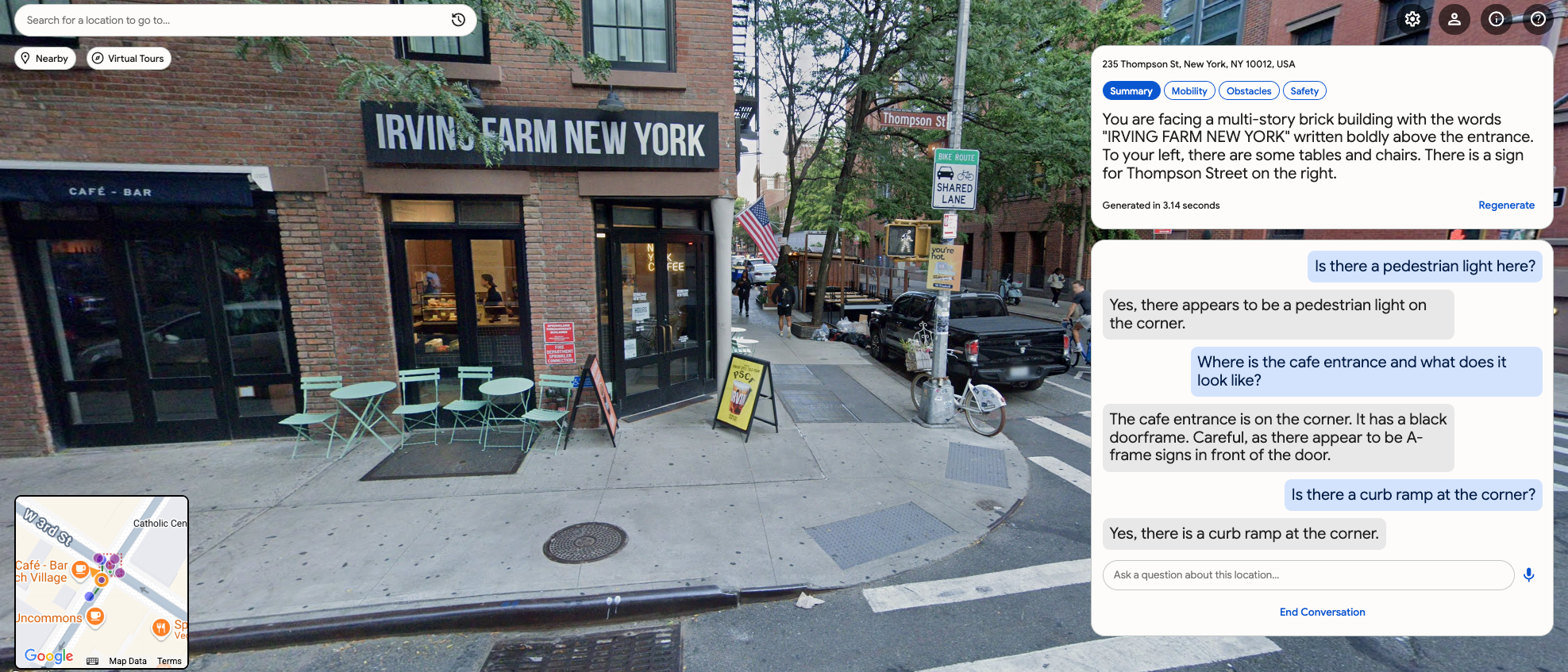}
  \vspace{-1.0em}
  \captionsetup{aboveskip=2pt}
  \captionof{figure}{We introduce our vision for \textit{\sysnames}—multimodal AI agents capable of understanding and responding to nuanced visual-spatial inquiries about the world by analyzing large-scale repositories of geospatial images combined with traditional GIS data sources. For example, \textit{StreetViewAI}~\cite{Froehlich_StreetViewAI_UIST2025} (above) makes street view accessible to blind users by combining geographic context, user information, and dynamic street view images into an MLLM, accessed via an AI chat interface and accessible screen reader controls.}
  \label{fig:teaser}
\end{strip}

\begin{abstract}
Interactive digital maps have revolutionized how people travel and learn about the world; however, they rely on pre-existing structured data in GIS databases (\textit{e.g.,} road networks, POI indices), limiting their ability to address geo-visual questions related to \textit{what} the world looks like. We introduce our vision for \textit{Geo-Visual Agents}—multimodal AI agents capable of understanding and responding to nuanced visual-spatial inquiries about the world by analyzing large-scale repositories of geospatial images, including streetscapes (e.g., Google Street View), place-based photos (e.g., TripAdvisor, Yelp), and aerial imagery (e.g., satellite photos) combined with traditional GIS data sources. We define our vision, describe sensing and interaction approaches, provide three exemplars, and enumerate key challenges and opportunities for future work.
\end{abstract}

\section{Introduction}
\label{sec:intro}
Over the last two decades, precise location sensing, pervasive internet connectivity, and interactive digital maps have transformed human mobility from travel planning to \textit{in situ} navigation.. Despite these advances, current mapping systems are confined to pre-existing structured geospatial data, leaving a vast repository of visual information---latent within street-level, aerial, and user-contributed imagery---untapped and inaccessible for answering what we term \textit{geo-visual questions}. That is, visually-oriented geographic questions about a location or route. Imagine, for example, a wheelchair user asking \myquote{Are there stairs leading up to the library on 35th?} or a blind traveler inquiring \myquote{Where is the door to the cafe and what does it look like?} 




In this workshop paper, we introduce our vision for \textit{\sysnames}---multimodal AI agents capable of understanding and responding to nuanced visual-spatial inquiries about the world by analyzing large-scale repositories of geospatial images (\textit{e.g.,} street-level and aerial imagery) combined with traditional GIS databases (\textit{e.g.,} road networks, POI databases, transit schedules). We envision \sysnames acting as \myquote{visual-spatial co-pilots} across a spectrum of contexts from \textit{a priori} travel planning to \textit{in situ} navigation. Crucially, while we expect many high-value user scenarios where a \sysname is actively sensing and processing visual-spatial data in real-time via AR glasses~\cite{Zhao_DesigningARVisToFacilitateStairNav_UIST2019, Fiannaca_Headlock_ASSETS2014, Lee_CookAR_UIST2024, Lee_GazePointAR_CHI2024} or smartphone cameras~\cite{Lo_NavigationARForVIP_Sensors2021, Yoon_LeveragingARForIndoorNav_ASSETS2019, Su_RASSAR_CHI2024}, an equally large set of questions can be answered by analyzing existing (and largely untapped) repositories of geo-related imagery---either on-demand (\textit{e.g.,} spinning up an AI agent to query and analyze sources) or via pre-computation. 




Our vision moves beyond the current paradigm of geospatial artificial intelligence (GeoAI)~\cite{Li_GeoAI_ISPRS2022, Janowicz_GeoAI_IJGIS2020, GoogleEarthAI_2025} such as \textit{CARTO AI}~\cite{CARTO_GenAI} and \textit{SuperMap}~\cite{SuperMap_AI_GIS}, which primarily focuses on large-scale data analysis for domain experts. Similarly, our work is related to but distinct from emerging paradigms in GIS research such as ``\textit{Autonomous GIS}''---AI-based scientific assistants that help \myquote{reason, derive, innovate, and advance geospatial solutions to pressing global challenges}~\cite{Li_AutonomousGIS_Arxiv2025}. Moreover, because our envisioned agents work primarily via multimodal conversational AI, we draw inspiration from recent work in \textit{Geospatial Visual Question Answering} (GVQA) such as \textit{MQVQA}~\cite{Zhang_MQVQA_IEEEGeoscience2023} and \textit{TAMMI}~\cite{Boussaid_TAMMI_CVPRWorkshop2025}, which attempt to imbue multimodal LLMs with domain-specific geographic knowledge; however, again these systems are aimed at analysts and function primarily on remote aerial imagery. While related, our focus is on addressing the personal, interactive, and often immediate needs of an individual planning travel or actively navigating a space. 




Below, we expand on our vision including a breakdown of visual-spatial inquiries, modalities of sensing and interaction, and three emerging examples, \textit{StreetViewAI}~\cite{Froehlich_StreetViewAI_UIST2025}, \textit{Accessibility Scout}~\cite{Huang_AccessibilityScout_UIST2025}, and \textit{BikeButler}. Throughout, we highlight key opportunities and open challenges.

\section{Geo-Visual Queries Across Travel Stages}
\label{sec:inquiry_taxonomy}
We envision \sysnames providing value across the full mobility cycle from pre-travel planning to \textit{in-situ} navigation. Below, we enumerate four travel stages and opportunities for \sysnames therein, focusing on accessibility but also broader user scenarios such as driving and biking. Selecting and fusing data sources will be a function of user task and data availability. For example, pre-travel planning may rely on streetscape images, user-contributed photos, and place-based reviews while in-situ navigation might combine these sources with visual content from a user's real-time camera feed (\textit{e.g.,} from AR glasses) and context sensing (\textit{e.g.,} travel mode inference, location). 




\textbf{Pre-travel planning.} In this phase, the user is not physically present at a location but planning a future visit. The agent acts as a remote, interactive guide, enabling detailed investigation and reducing uncertainty before travel. For example: (1) a blind parent planning a trip to a park may ask, \myquote{What kind of equipment does the playground have, and does it seem safe?} (2) A person with a mobility disability virtually investigates a route and inquires \myquote{Are there accessible curb ramps all the way to my doctor's office?} (3) A potential homebuyer may ask neighborhood-related questions such as \myquote{What do the streets look like?}, \myquote{Are there tree-lined sidewalks?}, and \myquote{How much graffiti is there?}

\textbf{While navigating.} During travel itself, the user is under cognitive and physical load, navigating their environment, making route choices, and dynamically avoiding obstacles. Here, the agent provides forward-looking information about the destination or upcoming maneuvers, enhancing situational awareness and facilitating \textit{in situ} travel decisions. For example: (1) A driver approaching an intersection asks, \myquote{You said to turn left at the next light. Are there any landmarks?} (2) A cyclist nearing a decision point queries, \myquote{Is there a protected bike lane at the next intersection, and which side of the road is it on?} (3) A rail user exiting a train asks, \myquote{Which exit is closest to the library's accessible entrance?}

\textbf{Destination arrival.} When arriving at a destination, the user is faced with a litany of ``last 10 meters'' problems related to the appearance of their destination, the path to and location of an entrance, and the presence of obstacles or safety issues. For example, (1) approaching their destination, a delivery driver may inquire \myquote{Where is the loading zone for this building?}; (2) a person meeting a friend in a busy plaza may ask, \myquote{I'm looking for the coffee shop; can you describe its storefront so I can more easily spot it?}. (3) a blind traveler's ride share arrives for pickup at a busy airport and asks, \myquote{Can you help me find the silver Toyota Camry with license plate KNI667?}.

\textbf{Indoor exploration.} Finally, upon entering a destination, the agent's role can shift to supporting micro-navigation through complex indoor environments like airports, stores, or office buildings. This stage presents a significant data challenge, as comprehensive visual and map datasets for indoor spaces are rare~\cite{Froehlich_GrandChallenges_Interactions2019}. For example, (1) a customer trying to find the location of a specific item in a hardware store may ask \myquote{Based on the aisle signs, which direction do I go to find the plumbing department?} (2) A low-vision traveler looking at an airport departure board: \myquote{Can you tell me which gate Delta Flight 850 is leaving from?}; (3) A wheelchair user in a large convention center: \myquote{Can you guide me to the nearest accessible restroom?}

Together, these scenarios illustrate how \sysnames can transform how we navigate and understand places, enhancing accessibility, offering landmark-based navigation, improving personal safety, and even leading to serendipitous discovery. Below, we describe potential data sources and then outline interaction modalities.

\section{Sensing and Data Sources}
The power of a \sysname lies in its ability to synthesize heterogeneous data sources, fusing visual evidence with structured geospatial data to form a holistic and accurate understanding of a place or route. We focus below on geo-related image sources rather than structured GIS data. 


\textbf{Streetscape Imagery.} Street view imagery (SVI)~\cite{Hou_ComprehensiveFrameworkForEvalSVI_IJAEOG2022, Li_SVI_Buildings2022}---such as \textit{Google Street View} (GSV), \textit{Cyclomedia}, \textit{KartaView}, and \textit{Mapillary}---provide a rich, large-scale image archive of the world. GSV alone has over 220 billion images spanning 10 million miles across 100 countries~\cite{Google_StreetView_Stats}. Such data can be used to analyze road conditions~\cite{Ali_PotholeDetectionViaGSV_ICOMP2024}, street markings (crosswalks~\cite{Li_MarkedCrosswalkAnalysisUsingGsv_EnvironmentAndPlanning2023, Ahmetovic_MiningGISImageryForCrosswalks_TACCESS2017}, bike lanes \cite{Rita_UsingDLandGSVforCyclistSafety_Sustainability2023}), sidewalk infrastructure (sidewalk material~\cite{Hosseini_CitySurfaces_SustainableCities2022}, curb ramps \cite{OMeara_RampNet_ICCV2025Workshop, Hara_TohmeCurbRampDetection_UIST2024}), bus stops~\cite{Kulkarni_BusStopCV_ASSETS23}, building facades~\cite{Kim_BarrierFreeEntranceDetectionInSVI_IEEEBigData2024}, graffiti \cite{Tokuda_QuantifyingGraffiti_IEEEBigComp2019}, trees and vegetation \cite{Li_QuantifyingShadeProvisionOfStreetTreesWithGSV_LandUP2018}, neighborhood health indicators \cite{Zou_DetectingAbandonedHousesFromGSV_ISPRS2021, wang_assessing_2024}, and more. Primary limitations include image recency~\cite{Wang_StreetViewForWhom_InSubmissions2025}, occlusions due to obstructing objects in front of the SVI camera (\textit{e.g.,} buses)~\cite{Saha_ProjectSidewalk_CHI2019}, and geographic distribution (images are distributed every 10-15 meters along roadways but not foot pathways or inside parks or buildings).


\textbf{User-Contributed Photos.} Place-based platforms like \textit{Google Places}, \textit{Yelp}, and \textit{TripAdvisor} contain vast, crowd-sourced libraries of photos tied to specific POIs, which provide a useful complement to SVI, including building interiors, curated (business uploaded) shots of storefronts, and pictures of menus, food~\cite{Gambetti_AIGenFoodReview_Arxiv2024}, and social activities (\textit{e.g.,} \cite{Zhang_ConsumerPostedYelpPhotosRestaurantSurvival_ManagementScience2023})---all which are often accompanied by user-contributed text (\textit{e.g.,} reviews). We found, however, that analysis of such multimodal data is less common in the literature. The key limitation here is data availability, particularly for unpopular or recently opened places, and social biases in \textit{who} uploads and \textit{why} (\textit{e.g.,} see ~\cite{Antoniou_MeasuresOfVgiQuality_ISPRS2015, Zhang_TheSpatialBiasOfVGI_AnnalsOfGis2018}).

\textbf{Aerial Imagery.} Aerial imagery from satellites, airplanes, or drones can provide high-resolution, top-down or oblique (45-degree angle) views of spatial structures, including building footprints, parking lots, vegetation, and pedestrian infrastructure~\cite{Hosseini_MappingTheWalk_CEUS2023}. While remote sensing and photogammetry research has existed for many decades---\textit{e.g.,} for land use classification, agriculture, disaster response, and military analyses~\cite{Janga_AReviewOfPracticalAIForRemoteSensing_RemoteSensing2023, Zhang_AIForRemoteSensing_IEEEGeoscience2022}---such techniques have not been applied to the \sysname context (\textit{e.g.,} answering end-user queries about parking lot locations, rooftop restaurant patios, or unmapped pedestrian shortcuts). Similar to streetscapes, aerial imagery can suffer from occlusions (from tree cover, clouds), shadows from tall buildings, and lack of availability. In the US, high-resolution aerial imagery is often provided by the federal government such as USGS~\cite{usgs_earthexplorer_2025} and NASA~\cite{nasa_landsat_2025}. 

\textbf{Robotic scans.} Robots such as autonomous vehicles, ground-based delivery robots, and drones~\cite{Song_AccessibleAreaMapper_SuMob2023, Su_FlyMeThrough_UIST2025} infused with sensor suites (cameras, LiDAR) can generate high-fidelity scans of the environment, producing not just images but 3D reconstructions with mensuration~\cite{Hu_UAVsAnd3DCityModeling_RemoteSensing2023}. While a potentially promising future data source, there is currently a lack of open data and APIs.


\textbf{Infrastructure-based Cameras.} Infrastructure-based cameras installed for traffic, weather, security, and safety monitoring provide real-time views of cities and uniquely offer dynamic information about pedestrian and car movement, human activity, weather conditions, and transient obstructions~\cite{Piadyk_StreetAware_Sensors2023, Rulff_StreetLevelMultimodalSensing_Arxiv2024, Jain_StreetNavLeveragingStreetCameras_UIST2024}; however, while some camera feeds are open---\textit{e.g.,} DOT traffic cameras---most are not and privacy is a key consideration. Moreover, there is a lack of density and availability (\textit{e.g.,} in rural areas).

\textbf{First-person Camera Streams.} Finally, first-person camera streams from AR glasses~\cite{Zhao_DesigningARVisToFacilitateStairNav_UIST2019, Fiannaca_Headlock_ASSETS2014, Lee_CookAR_UIST2024}, smartphone cameras~\cite{Lo_NavigationARForVIP_Sensors2021, Yoon_LeveragingARForIndoorNav_ASSETS2019, Su_RASSAR_CHI2024, Apple_Magnifier_Doors}, and dashcams~\cite{Park_MotivesOfDashCamVideoSharing_CHI2016, Zhanabatyrova_AutomaticMapUpdatesFromDashcam_IEEEIoT2023} are critical for in-situ travel stages, offering a real-time, egocentric view for navigation, identifying transient obstacles, and reading signs. While primarily used for immediate assistance, these streams could also help update or correct existing geospatial datasets in a continuous feedback loop (\textit{e.g.,} \cite{Zhanabatyrova_AutomaticMapUpdatesFromDashcam_IEEEIoT2023}). However, key considerations include high computational and power requirements, robust network connectivity, and privacy concerns for both the user and bystanders.
\section{Processing and Interpreting with AI}
Our vision relies not just on diverse forms of geospatial imagery and pre-existing GIS data but also advances in multimodal AI (\textit{e.g.,} scene understanding~\cite{Cordts_Cityscapes_CVPR16, Chen_DeepLab_TPAMI18}, object affordances~\cite{Hassanin_VisualAffordanceAndFunction_ACMComputSurvey2022, Lee_CookAR_UIST2024}, and spatial reasoning~\cite{Ranasinghe_SpatialReasoningInVLMs_CVPR2024, Chen_SpatialVLM_CVPR2024, Cheng_SpatialRGPT_Neurips2024, Fu_SceneLLM_Arxiv2024}) to extract semantic information and object relationships. While some analyses could be pre-computed for known high-value entities (\textit{e.g.,} presence and location of curb ramps~\cite{Hara_TohmeCurbRampDetection_UIST2024, OMeara_RampNet_ICCV2025Workshop}), we expect a long-tail of bespoke queries, which will require a \sysname to seek out, analyze, and synthesize image-based sources with pre-existing metadata in GIS databases in real-time.
\section{Delivering the Answers}
Finally, a crucial aspect of our vision is \textit{how} the agent delivers information, which is a function of the user's abilities, their current context, and the complexity and type of data. Regardless of delivery mode, agents need to report uncertainty and data provenance to build trust and mitigate error.


\textbf{Audio-First Interfaces:} For hands-free and/or eyes-free operation—essential for drivers, cyclists, and blind and low vision users—audio interfaces are critical (\textit{e.g.,} using earbuds or a smart speaker). The challenge, however, is providing well-structured verbal descriptions to convey complex visual information without overwhelming the user.


\textbf{Multimodal Interfaces:} Agents should also select and show relevant imagery. For instance, after describing an entrance, the agent could display a photo of the door (\textit{e.g.,} drawn from SVI or Yelp). The challenge lies in the AI’s ability to select the most appropriate photo(s)---appropriately cropped---from large archives.


\textbf{AI-Generated Abstracted Visualizations:} For highly complex spatial information, a raw photo or a long verbal description may be insufficient. An exciting frontier is the agent's ability to generate simplified, abstract diagrams on the fly---akin to a modern \textit{LineDrive} system~\cite{Agrawala_LineDrive_SIGGRAPH2001}. Making these abstractions accessible, perhaps tactilely, is also a critical area of open research.

\section{Case Study Applications}
\label{sec:case_studies}

To help showcase and concretize our vision, we highlight three emerging \sysname prototypes.


\textbf{StreetViewAI.} Current SVI tools are inaccessible to blind users. Our group is addressing this problem through the design of \textit{StreetViewAI}~\cite{Froehlich_StreetViewAI_UIST2025} (\autoref{fig:teaser}), which uses context-aware, real-time AI to support virtually exploring routes, inspecting destinations, or even remotely visiting tourist locations such as the Grand Canyon~\cite{Google_StreetView_GrandCanyon}. StreetViewAI provides accessible interactive controls for blind users to pan and move between panoramic images and dynamically converse with a live, multimodal AI agent about the scene and local geography. In a lab study, blind users effectively used StreetViewAI to virtually navigate streetscapes. Key challenges: reconciling users' mental models of SVI, a tendency to over-trust AI, and the difficulty of synthesizing rich visual data into concise audio.

\textit{\underline{AI Agent.}} StreetViewAI employs three separate AI subsystems. Most relevant is the \textit{AI Chat Agent}, which allows for conversational interactions about the user's current and past street views as well as nearby geography. The agent uses Google's \textit{Multimodal Live API}~\cite{google_multimodal_live_api}, which supports real-time interaction, function calling, and retains memory of all interactions within a single session. When the user initiates a chat either via typing or speaking, we transmit each GSV interaction along with the user's current view and geographic context (\textit{e.g.,} nearby places, current heading). Thus, users can ask about local geography, current and past views, and object relationships (\textit{e.g.,} \myquote{where is the entrance?}).



\textbf{Accessibility Scout.} Assessing the accessibility of unfamiliar environments is a critical but often laborious job for people with disabilities. While standardized checklists exist, they often fail to account for an individual’s unique and evolving needs. \textit{Accessibility Scout}~\cite{Huang_AccessibilityScout_UIST2025} is an LLM-based system designed to address this gap by generating personalized accessibility scans from images---\textit{e.g.,} from \textit{TripAdvisor}, \textit{Yelp}, and \textit{Airbnb}---to identify potential concerns based on self-reported abilities and interests. In user studies, we found that Accessibility Scout's personalized scans were more useful than generic ones and that its collaborative Human-AI approach was effective and built trust.

\textit{\underline{AI Agent.}} The Accessibility Scout pipeline begins by creating a structured user model in JSON format, initialized from a user's plain text description of their abilities and preferences. To assess an environment, the agent mimics how users assess environmental accessibility by first analyzing an image and the user's intent (\textit{e.g.}, \myquote{going on a date}) to identify potential tasks a user might perform, such as \myquote{dining} or \myquote{toileting}. The agent then decomposes these tasks into primitive motions like \myquote{grabbing} that are required to complete them. For each task, the agent analyzes the user model, task information, and segmented image to identify and describe environmental concerns. Crucially, the system is designed for Human-AI collaboration; users can provide feedback on identified concerns which the agent uses to update the user model.

\textbf{BikeButler.} Existing mapping tools define optimal bike routes using objective data like distance and elevation, but often ignore subjective qualities related to a cyclist's comfort and perceived safety. However, a desirable bike route depends on factors not found in standard GIS databases, such as the presence of tree-lined streets, pavement quality, or bike lane widths. \textit{BikeButler} is an early-stage prototype Geo-Visual Agent that generates personalized cycling routes by fusing structured data from \textit{OpenStreetMap} with visual analyses of SVI. The system creates routes optimized for a user’s specific profile (\textit{e.g.,} beginner, expert) and allows them to rate route segments, creating a feedback loop that refines their preferences for future journeys. 

\section{Discussion and Conclusion}
In this paper, we introduced our vision for \sysnames, dynamic and conversational AI co-pilots that can see and reason about the world in real-time. Our envisioned agents answer nuanced visual questions about the visual world---from a blind user navigating a complex intersection to a cyclist seeking the safest, most pleasant route. Our prototypes offer an initial window into this vision, offering personalized, interactive experiences extending far beyond current mapping services. 

Still, significant challenges remain, including: (1) \textit{Dynamic information synthesis}: creating agents that can intelligently select, fuse, and reason over a heterogeneous set of real-time and archived data sources; (2) \textit{Trust and transparency}: communicating uncertainty and data provenance; (3) \textit{Speech UIs}: effectively verbalizing complex visual information concisely via text or speech; (4) \textit{Personalization}: learning from a user's unique needs and preferences; (5) \textit{Spatial reasoning}: accurately tracking and modeling spatial relationships between objects and scenes; (6) \textit{Generative spatial abstractions}: dynamically generating spatial visualizations to help aid understanding. (7) \textit{Data source availability:} the availability of high-fidelity geospatial images both outdoors (\textit{e.g.,} streetscape images in parks, pedestrian-only pathways) and indoors (\textit{e.g.,} inside public buildings) as well as structured GIS data; (8) \textit{Data recency and correctness:} all techniques are reliant on up-to-date and accurate data. 

Addressing these challenges will require a concerted effort across disciplines from computer vision and HCI to accessibility and geospatial science. We look forward to discussing our \sysname vision at the ICCV workshop with the cross-disciplinary attendees.



{
    \small
    \bibliographystyle{ieeenat_fullname}
    \bibliography{main}
}

\end{document}